\documentclass{ifacconf}

\usepackage{siunitx}
\usepackage{mathtools}
\usepackage[ruled,vlined,algo2e]{algorithm2e}
\usepackage{verbatim}
\usepackage[version=4]{mhchem}
\usepackage[latin1]{inputenc}
\usepackage{tikz}
\usepackage{textcomp}
\usepackage{dirtytalk}
\usetikzlibrary{shapes,arrows}
\usepackage{enumerate}
\usepackage{color}
\usepackage{pgfplots}
\usetikzlibrary{calc}
\usepackage{mathrsfs}
\usepackage{upgreek}
\usepackage[outdir=./]{epstopdf}
\usepackage{lipsum}
\usepackage{amsbsy}
\usepackage{bm}
\usetikzlibrary{positioning}
\usetikzlibrary{shapes,arrows}
\usepackage{multirow}
\newcommand{\tx}{\textnormal}

\usepackage{subfigure} 
\usepackage{graphicx}

\begin{document}

\begin{frontmatter}

\title{Computationally Efficient Approach for Preheating of Battery Electric Vehicles before Fast Charging in Cold Climates}
\author{Ahad Hamednia, Jimmy Forsman, Nikolce Murgovski} \author{Viktor Larsson, Jonas Fredriksson}
\address{Department of Vehicle Energy and Motion Control, Volvo Car Corporation, Gothenburg 405 31, Sweden, (e-mail: ahad.hamednia@volvocars.com)}
\address{Department of Electrical Engineering, Chalmers University of Technology, 41296 Gothenburg, Sweden (e-mail: ahad.hamednia@chalmers.se).}

\begin{abstract}                          
This paper investigates battery preheating before fast charging, for a battery electric vehicle (BEV) driving in a cold climate. To prevent the battery from performance degradation at low temperatures, a thermal management (TM) system has been considered, including a high-voltage coolant heater (HVCH) for the battery and cabin compartment heating. Accordingly, an optimal control problem (OCP) has been formulated in the form of a nonlinear program (NLP), aiming at minimising the total energy consumption of the battery. The main focus here is to develop a computationally efficient approach, mimicking the optimal preheating behavior without a noticeable increase in the total energy consumption. The proposed algorithm is simple enough to be implemented in a low-level electronic control unit of the vehicle, by eliminating the need for solving the full NLP in the cost of only \SI{1}{Wh} increase in the total energy consumption.
\end{abstract}

\end{frontmatter}

\section{Introduction}
Lithium-ion (Li-ion) batteries have been recently a pre-dominant cell chemistry in the battery electric vehicle (BEV) market, due to their advantageous physical characteristics, such as high power and energy density, low self-discharge rate, long cycle life, and environmental friendliness~\cite{zhang22}. Nevertheless, the performance of Li-ion batteries is severely deteriorated at undesirably high and low temperatures. Particularly at low temperatures, the battery power is reduced significantly, resulting in reduced driving range and exacerbated range anxiety among BEV drivers~\cite{hao20}. Also, regenerative braking is limited or completely switched off in extremely cold weather, due to the phenomena of lithium plating that can strengthen potential safety hazards and decrease battery lifetime~\cite{waldmann18}. The reduced power availability at low battery temperatures is a major issue during fast charging, leading to significantly increased stop-over time, especially for BEVs with large batteries~\cite{zeng21}.

To address the aforementioned issues, different efforts have been made at the battery operational level, by developing an adequate battery thermal management (TM) system. A comprehensive review of previous research efforts on TM systems for Li-ion batteries is provided in~\cite{zichen21} and the references therein. Among the TM techniques, preheating the battery from very low temperatures to a desired high temperature, especially before departure and/or fast charging, is known as an effective approach~\cite{zhang17b,perez17,wang18,wu20,wang22}. The preheating have traditionally been performed through (1): internal heating scheme by applying a current to a battery and thus, generating a heat to warm up the battery because of the internal resistance; (2): external heating scheme by transferring the heat generated by an external component, e.g. high-voltage coolant heater (HVCH), via a medium, e.g. coolant, to the battery pack~\cite{peng19}. To preheat the battery before departure, several attempts have been conducted in the technical literature~\cite{ji13,damay13,zhu13}. It is also possible to apply various preheating strategies while driving, as presented in~\cite{zhang17,hamednia2022a,hamednia2022b}. Our earlier results in ~\cite{hamednia2022a} show that it is more energy-optimal to not preheat the battery at home, but rather only some time period prior to the arrival at the charging station, using HVCH at maximum power. Despite recent studies on the preheating before fast charging, to the best of our knowledge a computationally efficient algorithm has not been developed to be applicable on a real vehicle.

This paper addresses battery preheating of a BEV before reaching a planned fast-charging station, where the battery is soaked to cold ambient before the vehicle's departure. The main idea is to devise a simple approach that mimics the optimal behavior obtained from our previous investigations, while providing a proper trade-off between optimality and
computational burden. Such approach is much easier and cheaper to industrialize. Accordingly, the proposed strategy consists of a forward and a backward simulations of the vehicle system dynamics. In the forward simulation, the system dynamics are rolled out starting from a given battery temperature and state of charge (SoC), where the HVCH is not used for heating the battery throughout the vehicle's trip. Conversely in the backward simulation, the reverted system dynamics are simulated backwards, starting from a target battery temperature and SoC, where the HVCH is used to heat the battery, at maximum power. The backward simulation continues until to a step, in which the resultant battery temperature value from the backward simulation is equal the one belonging to the forward simulation.

The remainder of this paper is organized as follows: the system modelling including the vehicle powertrain, and electrical and thermal modellings are given in Section~\ref{sec:model}. Section~\ref{sec:method}, corresponds to the OCP formulation and developing the computationally efficient algorithm. In Section~\ref{sec:res}, simulation results are presented. Finally, Section~\ref{sec:con} concludes the paper.

\begin{figure}[t!]
 \centering
 \includegraphics[width=.875\linewidth]{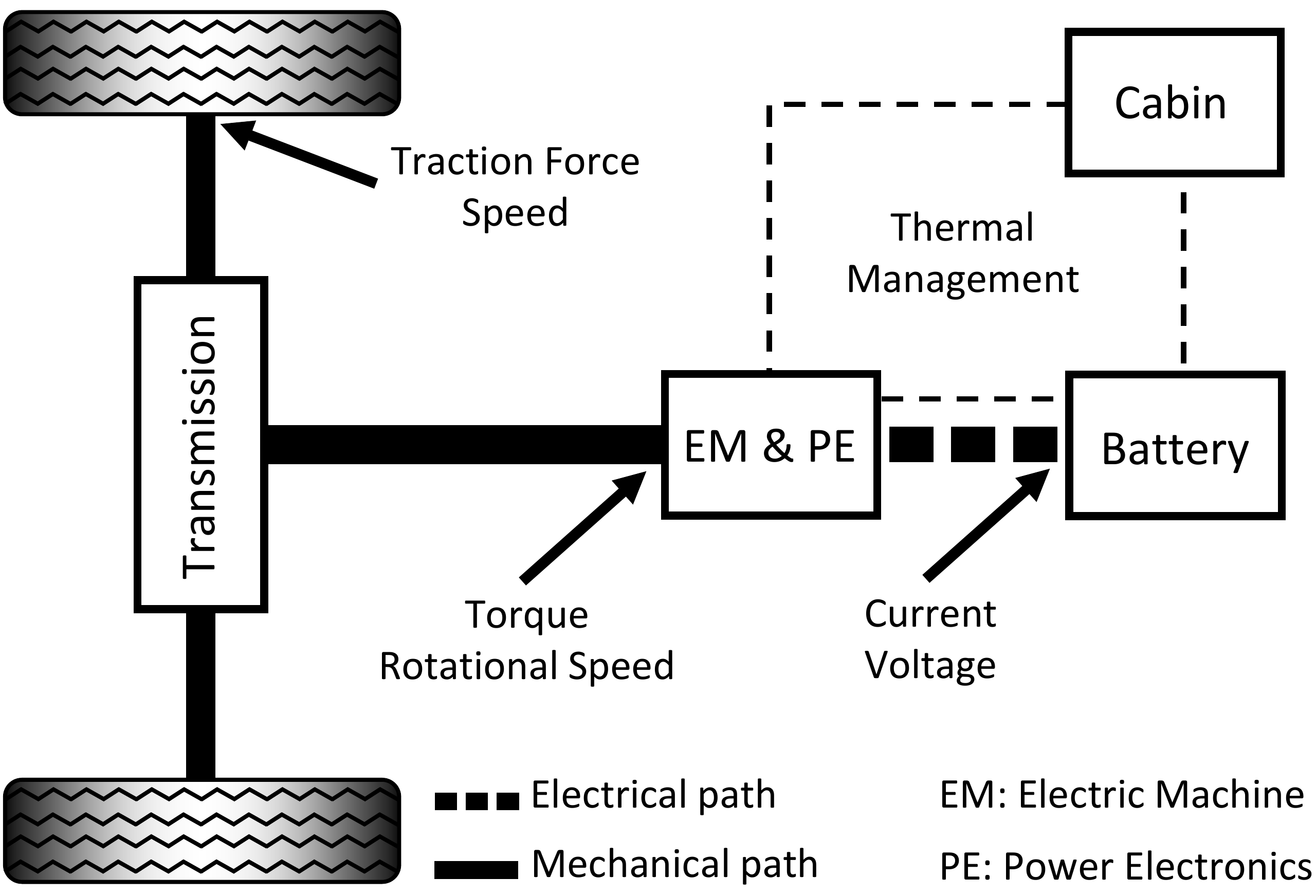}\vspace{-.1cm}
  \caption{\footnotesize Schematic diagram of the studied electric powertrain, including battery, electric machine, power electronic devices, transmission system, and thermal management system.}
  \label{fig:pow_sketch}
\end{figure}

\section{Modelling}\label{sec:model}
This section addresses the modelling of a BEV. Primarily, a brief overview of key BEV powertrain components is given. Later, electrical and thermal governing dynamics of the powertrain are described.

\subsection{Vehicle Powertrain}\label{subsec:pow}
As depicted in Fig.~\ref{fig:pow_sketch}, the studied powertrain consists of battery, electric machine (EM), power electronic (PE) devices, transmission system, and thermal management system. Depending on operating mode of the EM, the electric power flow between the battery and EM is bidirectional. Accordingly, the electrical energy from the EM is stored in the battery during the EM's generating mode. On the other hand, the EM when operated in motoring mode receives electrical power from the battery and provides propulsion power via the transmission system to the wheels through a mechanical path. Thus, the EM torque and rotational speed are respectively translated to traction force and vehicle speed. 

In addition to the traction force, multiple forces are applied to a longitudinal motion of the vehicle, such as aerodynamic drag, rolling resistance, and gravitational load. Thus, the net force exerted on the vehicle can be obtained for a given set of the aforementioned forces, using Newton's second law of motion. An example of formulating the longitudinal dynamics is carried out in~\cite{hamednia2022a}. In this paper, we assume that the propulsion power demand for the entire driving cycle is given as a trajectory input over a number of segments, instead of modelling the longitudinal dynamics. Likewise, the vehicle speed is supplied for each segment of the drive cycle, i.e. it is assumed that these trajectories are provided by the trip planning functionality of the on-board navigation system.


\begin{figure*}[t!]
\centering
\subfigure[Battery discharge power limit.]{
 \includegraphics[width=.425\linewidth]{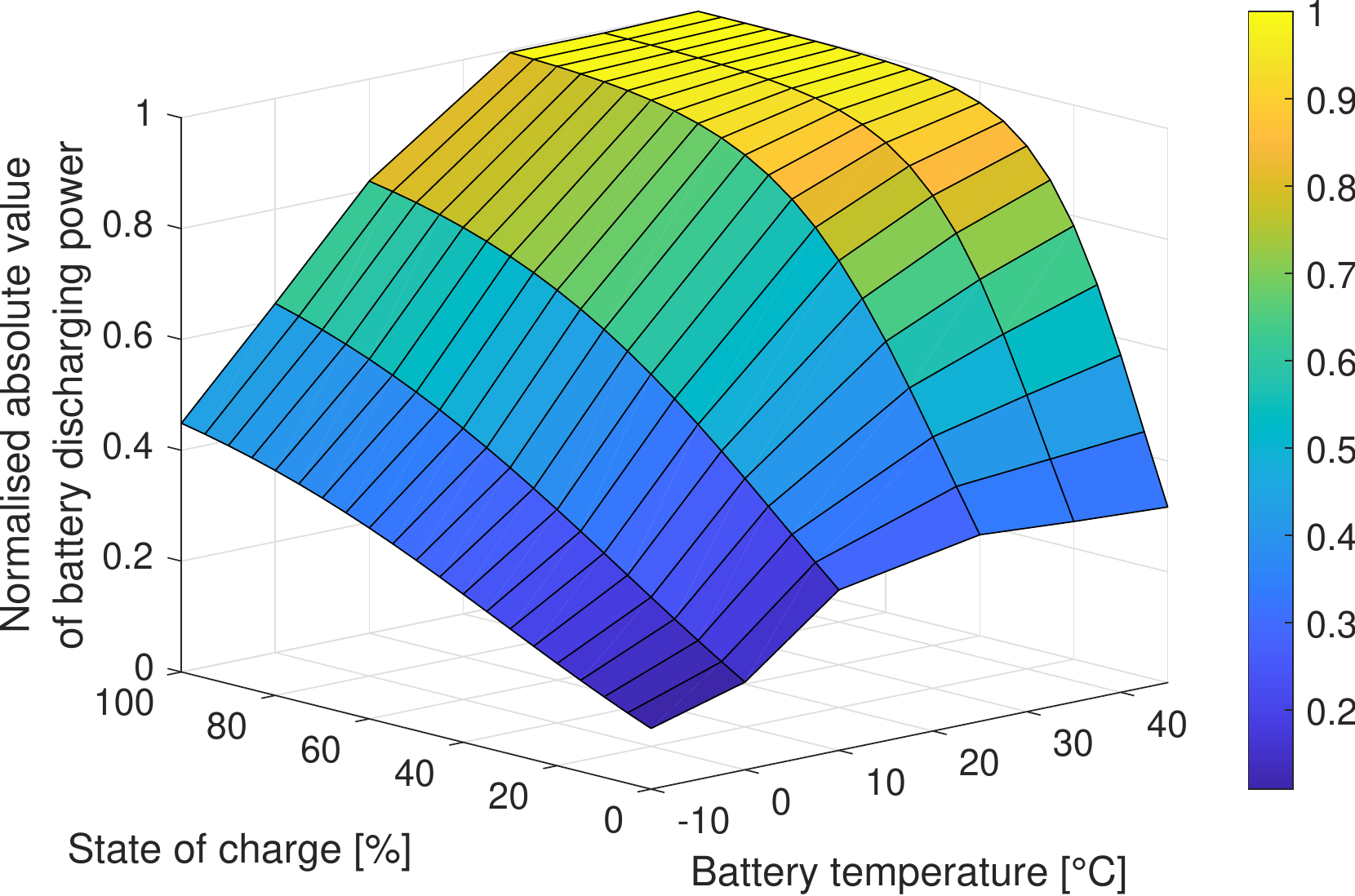}\hspace{1cm}
\label{fig:pbdchglim}
}
\subfigure[Battery charge power limit.]{

 \includegraphics[width=.425\linewidth]{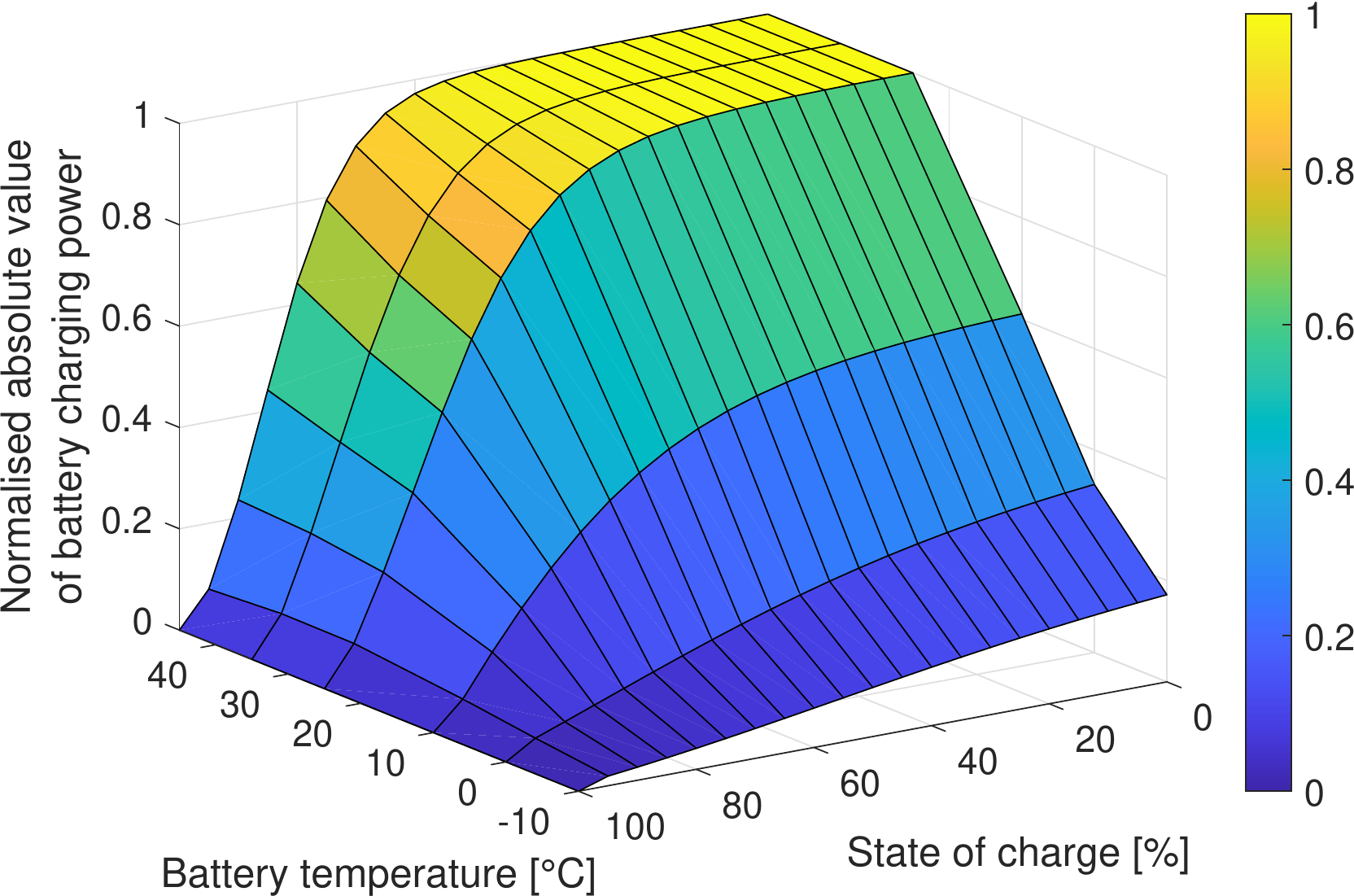}
\label{fig:pbchglim}
}
\caption{Normalised absolute value of battery discharge and charge power limits for a given pair of battery temperature and SoC. These figures are taken from~\cite{hamednia2022a}.}\vspace{-.1cm}
\label{fig:pblim}
\end{figure*}
\subsection{Electrical Modelling}\label{subsec:el}
In order to describe how the battery SoC changes over time, an equivalent circuit in 
is used for electrical modelling of the battery pack. The circuit consists of an open-circuit voltage $U_\tx{oc}$ and an internal resistance $R$. The open-circuit voltage is generally a nonlinear increasing function of SoC, obtained via offline experiments at various ambient temperatures and battery aging stages. Also, the internal resistance is usually proportional to the inverse of battery temperature~\cite{zhu18}. The battery SoC dynamics is given by
\begin{align}
    \dot{\tx{soc}}(t)=-\frac{P_\tx{b}(t)}{C_\tx{b}U_\tx{oc}(\tx{soc})},
    \label{eq:soc_dyn}
\end{align}
where $t$ is trip time, $P_\tx{b}$ is battery power including internal resistive losses, and $C_\tx{b}$ is maximum battery capacity. According to \eqref{eq:soc_dyn}, $P_\tx{b}$ is negative when charging, and is positive while discharging. Also, $P_\tx{t}$ is battery power after internal resistive losses.

The battery power available while discharging and charging is limited as functions of battery temperature and SoC as
\begin{align}
P_\tx{b}(t)\in 
	[P_\tx{b,chg}^{\min}(\tx{soc},T_\tx{b}),P_\tx{b,dchg}^{\max}(\tx{soc},T_\tx{b})],
\label{eq:pblim} 
\end{align}
where $P_\tx{b,dchg}^{\max}>0$ is maximum battery power (discharging) and $P_\tx{b,chg}^{\min}<0$ is minimum battery power (charging). Note that $\tx{soc}$ and $T_\tx{b}$ are functions of trip time in \eqref{eq:soc_dyn}, \eqref{eq:pblim}, and the equations hereafter. However, the explicit dependence is not shown for brevity. The power limits demonstrated in
Fig.~\ref{fig:pblim} are representative, but generic data, from a vehicle original equipment manufacturer (OEM). According to Fig.~\ref{fig:pblim}, maximum battery power (discharging) peaks at high SoC and battery temperature, and drops as they decrease. Discharge power limit is of significant interest while driving as it can directly limit the provided propulsion power. On the contrary, the highest absolute power (charging) is observed when SoC is low while battery temperature is high, and drops as SoC increases and battery temperature decreases. Thus, for a cold battery it is of interest to warm up the battery pack to a suitable temperature prior to charging, in order to allow high-power charging, thereby reducing the charging time.

\subsection{Thermal Modelling}\label{subsec:th}
\begin{figure}[t!]
 \centering
 \includegraphics[width=.875\linewidth]{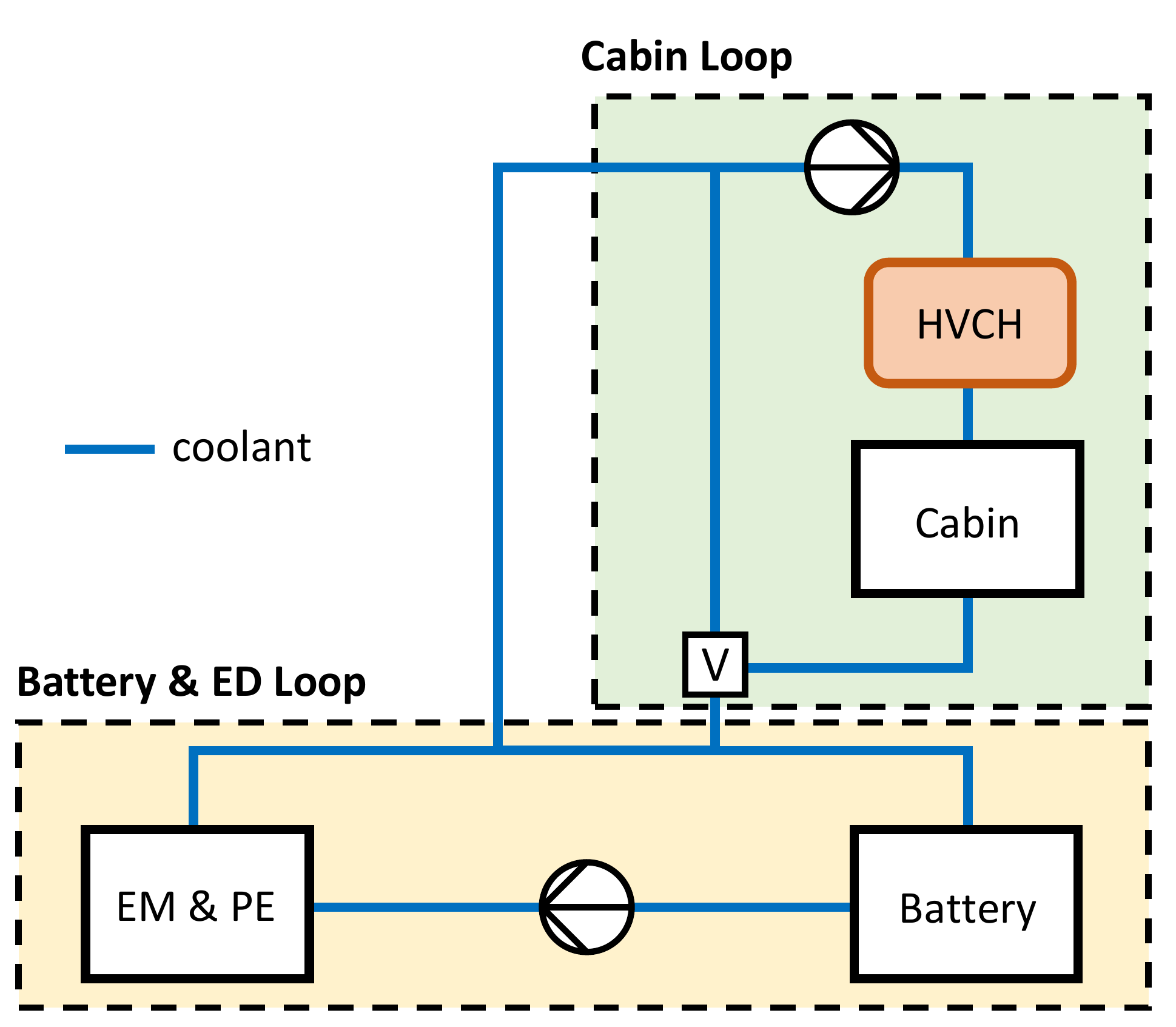}
  \caption{\footnotesize Schematic diagram of thermal system, consisting of two thermal loops, i.e. battery and ED loop and cabin loop. Heat exchange between the two loops is facilitated by the valve V.}\vspace{-.1cm}
  \label{fig:th}
\end{figure}

The thermal management system in question is illustrated in Fig.~\ref{fig:th}, including two thermal loops, i.e. battery and electric drivetrain (ED) loop and cabin loop. PE devices and EM are the two major heat generating components within the ED. The two loops represent two thermal subsystems with different heat sinks and sources. Heat transfer between the loops is performed through a coolant circulated by multiple pumps. The consumed power by the pumps is included in the auxiliary power consumption. For cold ambient operation, the heat from the ED components is assumed to always be available for heating of the battery. The HVCH is also used for heating cabin and/or battery. Depending on the desired amount of cabin and battery heating, the valve V is adjusted to accommodate only cabin heating, only battery heating, or their mix. Note that in warm ambient operation, the heat from the ED components may cause excessive heating of the battery, beyond operating limits. Thus, it is most likely desirable to separate the ED components from the battery thermal loop, in order to cool the battery, using air conditioning system and/or heat pump. The ED components can also be cooled by a radiator.

According to the fundamental thermodynamic principle, the battery pack's thermal dynamics is described as
\begin{align}
\begin{split}
&\dot{T_\tx{b}}(t)=\frac{1}{c_\tx{p}m_\tx{b}}\Big(\eta_\tx{hvch}P^\tx{b}_\tx{hvch}(t)+\gamma(t)(T_\tx{amb}(t)-T_\tx{b}(t))\\
&\hspace{4cm}+Q_\tx{Joule}(t)+Q_\tx{ed}(t)\Big),
\end{split}\label{eq:tb_dyn}
\end{align}
where $c_\tx{p}$ is specific heat capacity of the battery pack, $m_\tx{b}$ is total battery mass, $P^\tx{b}_\tx{hvch}$ is HVCH power converted with the efficiency of $\eta_\tx{hvch}$ for heating the battery pack, $\gamma$ is parasitic coefficient of heat transfer between the battery and the ambient, $T_\tx{amb}$ is ambient temperature, and $Q_\tx{Joule}$ is irreversible ohmic Joule heat induced by the battery internal resistive losses given by
\begin{align}
    &Q_\tx{Joule}(t)=R(T_\tx{b})\frac{P_\tx{b}^2(t)}{U_\tx{oc}^2(\tx{soc})}.\label{eq:joule}
\end{align}
Furthermore, $Q_\tx{ed}$ is the heat generated from ED power losses defined as
\begin{align}
    &Q_\tx{ed}(t)=\eta_\tx{ed}^\tx{Q}\big(1-\eta_\tx{ed}^\tx{e}(v,F)\big)P_\tx{prop}(t),\label{eq:ed}
\end{align}
where $\eta_\tx{ed}^\tx{Q}$ is the efficiency of ED power loss conversion to thermal power for heating the battery, $\eta_\tx{ed}^\tx{e}$ is the EM's lumped efficiency, which is dependent on the vehicle speed $v$ and traction force $F$. Also, $P_\tx{prop}$ is propulsion power including the internal losses of the powertrain.

\section{Battery Preheating before Fast Charging}\label{sec:method}
\begin{figure}[t!]
 \centering
 \includegraphics[width=.95\linewidth]{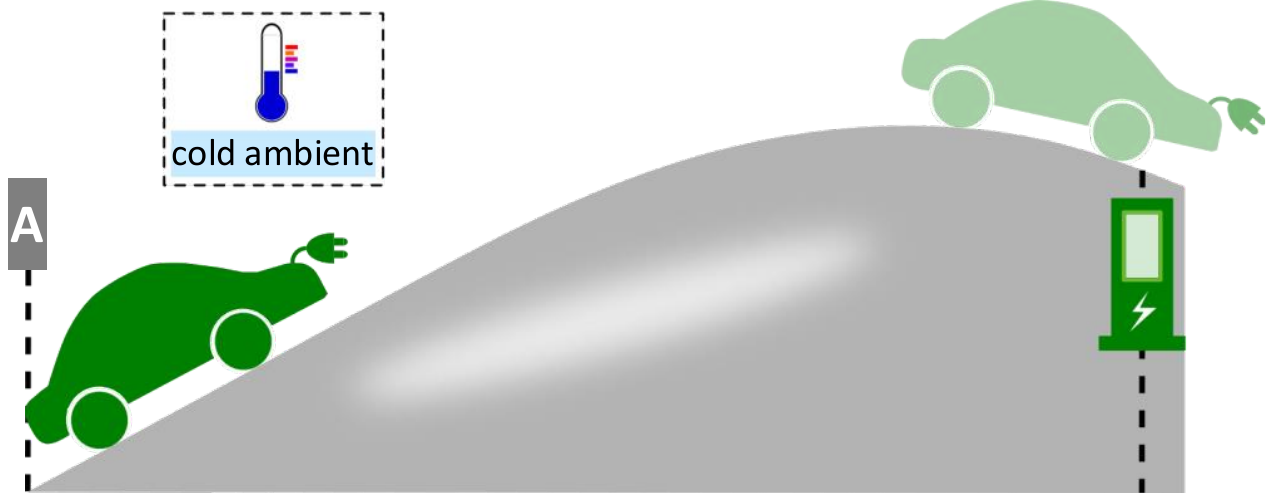}
  \caption{\footnotesize Studied scenario; a BEV is driving until reaching a planned fast-charging occasion. The vehicle starts its mission from point \textbf{A} with a battery soaked to cold ambient before the vehicle's departure.}\vspace{-.1cm}
  \label{fig:scenario}
\end{figure}

Consider a BEV driving until it reaches a planned fast-charging station, as in Fig.~\ref{fig:scenario}, where the trip starts from point \textbf{A} with a fully-charged battery. The battery is soaked to cold ambient before the vehicle's departure. The battery temperature is changed by the aforementioned heating sources within the powertrain. Also, it is assumed that the cabin heat demand is always fulfilled by HVCH throughout the trip. 

\subsection{Optimal Control Problem Formulation}\label{subsec:pf}
In the following, an optimisation problem is formulated to achieve optimal battery preheating before fast charging, by minimising the total energy consumption of the vehicle during the entire mission, as
{\allowdisplaybreaks
\begin{subequations} \label{eq:p1}
\begin{align}
&\min_{P^\tx{b}_{\tx{hvch}}, P_\tx{b}} \int_{t_0}^{t_\tx{f}}P_\tx{b}(t)\tx{d}t\\
& \text{subject to:} \nonumber\\
&\dot{\tx{soc}}(t)=-\frac{P_\tx{b}(t)}{C_\tx{b}U_\tx{oc}(\tx{soc})}\label{eq:p1_soc}\\
\begin{split}
&\dot{T_\tx{b}}(t)=\frac{1}{c_\tx{p}m_\tx{b}}\Big(\eta_\tx{hvch}P^\tx{b}_\tx{hvch}(t)+\gamma(t)(T_\tx{amb}(t)-T_\tx{b}(t))\\
&\hspace{4cm}+Q_\tx{Joule}(t)+Q_\tx{ed}(t)\Big)
\end{split} \label{eq:p1_Tb}\\
\begin{split}
&P_\tx{b}(t)=R(T_\tx{b})\frac{P_\tx{b}^2(t)}{U_{\tx{oc}}^2(\tx{soc})}+P^\tx{b}_\tx{hvch}(t)+P^\tx{c}_\tx{hvch}(t)\\
&\hspace{4cm}+P_\tx{aux}(t)+P_\tx{prop}(t)  
\end{split}\label{eq:p1_Pb}\\
&\tx{soc}(t) \in [\tx{soc}_{\min},\tx{soc}_{\max}] \label{eq:p1_socbound}\\
&T_\tx{b}(t) \in [T_\tx{b}^{\min}(t),T_\tx{b}^{\max}(t)] \label{eq:p1_Tbbound}\\
&P^\tx{b}_\tx{hvch}(t) \in [0,P^{\max}_{\tx{hvch}}-P^\tx{c}_\tx{hvch}(t)] \label{eq:p1_Pbhvchbound}\\
&P_\tx{b}(t) \in [P_\tx{b,chg}^{\min}(\tx{soc},T_\tx{b}),P_\tx{b,dchg}^{\max}(\tx{soc},T_\tx{b})]\label{eq:p1_Pbbound}\\
&T_\tx{b}(t_0)=T_\tx{b0}, \quad \tx{soc}(t_0)=\tx{soc}_0 \label{eq:p1_int}\\
&T_\tx{b}(t_\tx{f})\geq T_\tx{bf}, \quad \tx{soc}(t_\tx{f})\geq \tx{soc}_\tx{f} \label{eq:p1_final}
\end{align}
\end{subequations}}%
where $t_0$ and $t_\tx{f}$ are initial and final trip times, respectively, $P_\tx{aux}$ is auxiliary load demand, $P^\tx{c}_\tx{hvch}$ is ambient dependent HVCH power demand used for heating cabin, $P^{\max}_{\tx{hvch}}$ and is the maximum deliverable HVCH power, $\tx{soc}_{\min}$ and $\tx{soc}_{\max}$ are the bounds on the battery SoC, $T_\tx{b}^{\min}$ and $T_\tx{b}^{\max}$ are the battery temperature limits, $T_\tx{b0}$ and $T_\tx{bf}$ are initial and final battery temperatures, respectively, $\tx{soc}_0$ and $\tx{soc}_\tx{f}$ are initial and final SoC, respectively. 

\subsection{Computationally Efficient Algorithm}\label{subsec:cea}
The problem \eqref{eq:p1} is a nonlinear program (NLP) with soc and $T_\tx{b}$ defined as the state variables, and $P^\tx{b}_\tx{hvch}$ and $P_\tx{b}$ defined as the control input and measured output, respectively. For the purpose of an offline analysis, the NLP can be solved on a PC, by commonly used nonlinear optimisation tools, e.g. CasADi~\cite{andersson19}. However, in order to solve problem \eqref{eq:p1} online in a vehicle, it is essential to significantly reduce the computational burden. To do so, the knowledge from the offline analysis can preferably be utilized. According to our observations, it is usually optimal to use HVCH at its maximum available power just some period before reaching the fast charging station, in order to take the battery temperature to the desirable value, e.g. \SI{25}{^\circ C}, at the arrival of charging station. By such an approach, the energy waste in the form of the thermal leakage to ambient from early battery heating is avoided.

\begin{figure}[t!]
 \centering
 \includegraphics[width=.85\linewidth]{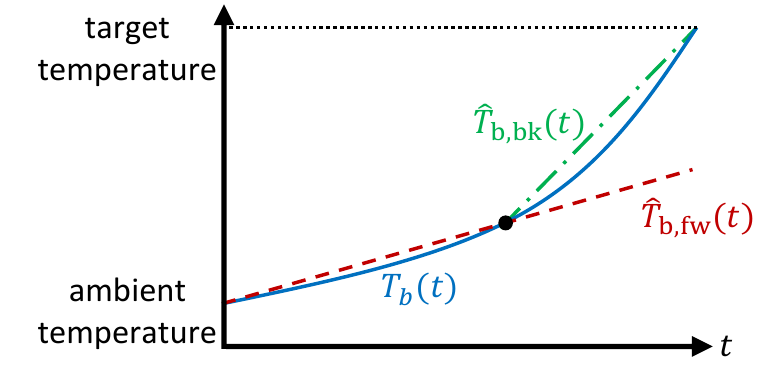}
  \caption{\footnotesize Battery temperature versus time; $T_\tx{b}$ is an example trajectory of battery temperature (obtained by solving problem \eqref{eq:p1}), $\hat{T}_\tx{b,fw}$ and $\hat{T}_\tx{b,bk}$ denote the battery temperature derived from forward and backward simulations, respectively.}\vspace{-.1cm}
  \label{fig:fw_bk}
\end{figure}

To reduce the computational complexity, we propose a heuristic approach, which consists of forward and backward simulations of the system dynamics. To do so, primarily the system dynamics \eqref{eq:soc_dyn} and \eqref{eq:tb_dyn} are discretized, using the first-order Euler method. In the forward simulation, given in \textbf{Algorithm 1}, the system dynamics are rolled out towards the end, starting from $\tx{soc}_0$ and $T_\tx{b0}$, where the HVCH power demand for heating the battery is always zero. Later in \textbf{Algorithm 2}, the system dynamics are reverted and simulated backwards, starting from $\tx{soc}_\tx{f}$ and $T_\tx{bf}$, where the HVCH power demand for heating the battery is set to the maximum available amount. In the remainder of the paper, the variables notated with subscripts/superscripts `fw' or `bk', represent the previously introduced variables that now belong specifically to the forward or backward simulation, respectively. Furthermore, the variables obtained from applying the heuristic algorithm are displayed with the hat ( $\hat{}$ ) symbol. Note that in the backward simulation, we use the SoC and battery temperature at instant $k+1$, when calculating the battery current at instant $k$. This is a reasonable approximation if the sampling interval is short enough, as the SoC and battery temperature only change very slightly between two consecutive instants due to their large time constants. The backward simulation continues until to a time step $k$, in which $\hat{T}_\tx{b,bk}^{(k)}=\hat{T}_\tx{b,fw}^{(k)}$, as shown in Fig.~\ref{fig:fw_bk}. The estimated battery temperature $\hat{T}_\tx{b}$ and SoC $\hat{\tx{soc}}$ for the whole mission are obtained, respectively as
\begin{align*}
&\hat{T}_\tx{b}=[\hat{T}_\tx{b,fw}^{(1:k)},\hat{T}_\tx{b,bk}^{(k+1:N+1)}],\\
&\hat{\tx{soc}}=[\hat{\tx{soc}}_\tx{fw}^{(1:k)},\hat{\tx{soc}}_\tx{bk}^{(k+1:N+1)}],
\end{align*}
where $N+1$ is the number of state samples.

\begin{algorithm2e}
\SetAlgoLined
$\hat{\tx{soc}}_\tx{fw}^{(1)}=\tx{soc}_0\in[\tx{soc}_{\min},\tx{soc}_{\max}]$\\
\vspace{0.1cm}
$\hat{T}_\tx{b,fw}^{(1)}=T_\tx{b0}\in[T_\tx{b}^{\min},T_\tx{b}^{\max}]$\\
\vspace{0.1cm}
\For{$k=1,\dots,N$}{
$\hat{P}_\tx{t,fw}^{(k)}=P_\tx{aux}^{(k)}+P_\tx{hvch}^{c}{^{(k)}}+P_\tx{prop}^{(k)}$\\
\vspace{0.1cm}
$\hat{I}_\tx{b,fw}^{(k)}=\frac{\hat{U}_\tx{oc,fw}(\hat{\tx{soc}}_\tx{fw}^{(k)})-\sqrt{\hat{U}^2_\tx{oc,fw}(\hat{\tx{soc}}_\tx{fw}^{(k)})-4\hat{R}_\tx{fw}(\hat{T}_\tx{b,fw}^{(k)})\hat{P}_\tx{t,fw}^{(k)}}}{2\hat{R}_\tx{fw}(\hat{T}_\tx{b,fw}^{(k)})}$\\
\vspace{0.1cm}
$\hat{\tx{soc}}_\tx{fw}^{(k+1)}=-\frac{\Delta t\hat{I}_\tx{b,fw}^{(k)}}{C_\tx{b}}+\hat{\tx{soc}}_\tx{fw}^{(k)}$\\
\vspace{0.1cm}
$\hat{T}_\tx{b,fw}^{(k+1)}=\frac{\Delta t}{c_\tx{p}m_\tx{b}}\Big(\gamma^{(k)}(T_\tx{amb}^{(k)}-\hat{T}_\tx{b,fw}^{(k)})$\\
\vspace{0.1cm}
$\hspace{2.25cm}+\hat{R}_\tx{fw}(\hat{T}_\tx{b,fw}^{(k)})\hat{I}_\tx{b,fw}{^{(k)}}^2+Q_\tx{ed}^{(k)}\Big)+\hat{T}_\tx{b,fw}^{(k)}$\\
\vspace{0.1cm}
$\hat{P}_\tx{b,fw}^{(k)}=\hat{P}_\tx{t,fw}^{(k)}+\hat{R}_\tx{fw}(\hat{T}_\tx{b,fw}^{(k)})\hat{I}_\tx{b,fw}{^{(k)}}^2$
}
\caption{Forward simulation} \label{al:prefw}
\end{algorithm2e}

\begin{algorithm2e}
\SetAlgoLined
$\hat{\tx{soc}}_\tx{bk}^{(N+1)}=\tx{soc}_\tx{f}\in[\tx{soc}_{\min},\tx{soc}_{\max}]$\\
\vspace{0.1cm}
$\hat{T}_\tx{b,bk}^{(N+1)}=T_\tx{bf}\in[T_\tx{b}^{\min},T_\tx{b}^{\max}]$\\
\vspace{0.1cm}
\For{$k=N,\dots,1$}{
$\hat{P}_\tx{t,bk}^{(k)}=P_\tx{aux}^{(k)}+P_\tx{hvch}^{c}{^{(k)}}+\hat{P}_\tx{hvch}^{b}{^{(k)}}+P_\tx{prop}^{(k)}$\\
\vspace{0.1cm}
$\hat{I}_\tx{b,bk}^{(k)}=$\\
\vspace{0.1cm}
$\hspace{1cm}\frac{\hat{U}_\tx{oc,bk}(\hat{\tx{soc}}_\tx{bk}^{(k+1)})-\sqrt{\hat{U}^2_\tx{oc,bk}(\hat{\tx{soc}}_\tx{bk}^{(k+1)})-4\hat{R}_\tx{bk}(\hat{T}_\tx{b,bk}^{(k+1)})\hat{P}_\tx{t,bk}^{(k)}}}{2\hat{R}_\tx{bk}(\hat{T}_\tx{b,bk}^{(k+1)})}$\\
\vspace{0.1cm}
$\hat{\tx{soc}}_\tx{bk}^{(k)}=\frac{\Delta t\hat{I}_\tx{b,bk}^{(k)}}{C_\tx{b}}+\hat{\tx{soc}}_\tx{bk}^{(k+1)}$\\
\vspace{0.1cm}
$\hat{T}_\tx{b,bk}^{(k)}=f(\hat{T}_\tx{b,bk}^{(k+1)},\hat{\tx{soc}}_\tx{bk}^{(k+1)})$\\
\vspace{0.1cm}
$\hat{P}_\tx{b,bk}^{(k)}=\hat{P}_\tx{t,bk}^{(k)}+\hat{R}_\tx{bk}(\hat{T}_\tx{b,bk}^{(k)})\hat{I}_\tx{b,bk}{^{(k)}}^2$\\
\vspace{0.2cm}
\textbf{if} $\hat{T}_\tx{b,bk}^{(k)}==\hat{T}_\tx{b,fw}^{(k)}$\\
\vspace{0.1cm}
\hspace{.5cm}$\hat{T}_\tx{b}=[\hat{T}_\tx{b,fw}^{(1:k)},\hat{T}_\tx{b,bk}^{(k+1:N+1)}]$\\
\vspace{0.1cm}
\hspace{.5cm}\text{stop the for loop}\\
\vspace{0.1cm}
\textbf{end}
\vspace{0.1cm}
}
\caption{Backward simulation} \label{al:prebk}
\end{algorithm2e}

\section{Results}\label{sec:res}
In this section, the performance of the heuristic proposed algorithm is evaluated, by comparing the heuristic solution with the solution obtained from solving the NLP \eqref{eq:p1}. 

The simulations are performed for a BEV that starts its trip with a battery soaked to cold ambient at $\SI{-7}{^\circ C}$. The cabin compartment heating demand is considered as a fixed value, given a constant low ambient temperature throughout the vehicle's trip. Also, the vehicle speed profile for \SI{60}{min} (\SI{73}{km}) of the vehicle's drive has been provided based on real-world measurements. We assume a planned fast-charging station at the $60^\tx{th}\,$km. The vehicle and simulation parameters are provided in Table \ref{tab:par}. The NLP \eqref{eq:p1} is discretized using the Runge-Kutta $4^\tx{th}$ order method~\cite{butcher76}, with a sampling interval of \SI{30}{s}. The discretized problem is solved in Matlab with the solver IPOPT, using CasADi~\cite{andersson19}.

\subsection{NLP Solution versus Heuristic Solution}\label{subsec:comp}
\begin{figure}[t!]
\centering
\subfigure[Battery temperature, target temperature, and ambient temperature.]{
 \includegraphics[width=.9\linewidth]{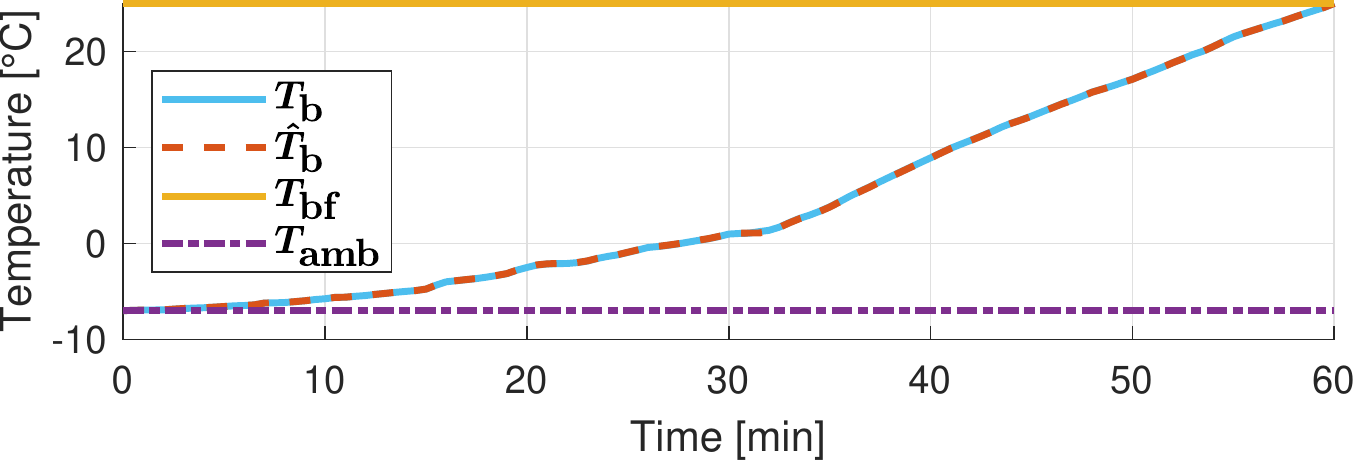}
\label{fig:tb}
}
\subfigure[HVCH power for battery heating.]{
 \includegraphics[width=.9\linewidth]{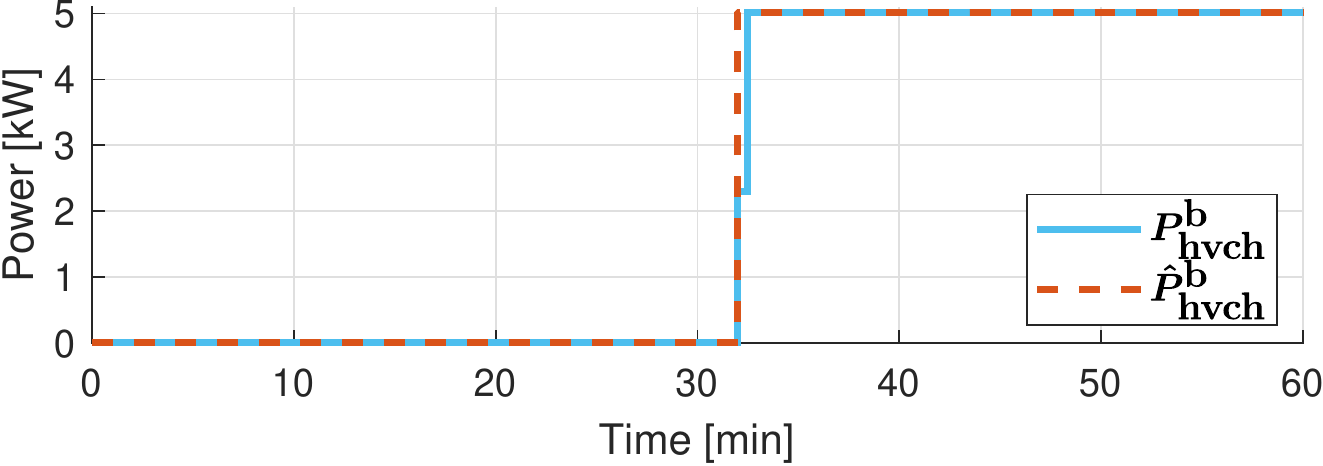}
\label{fig:hvch}
}
\subfigure[Joule heating and ED circuit losses .]{
 \includegraphics[width=.9\linewidth]{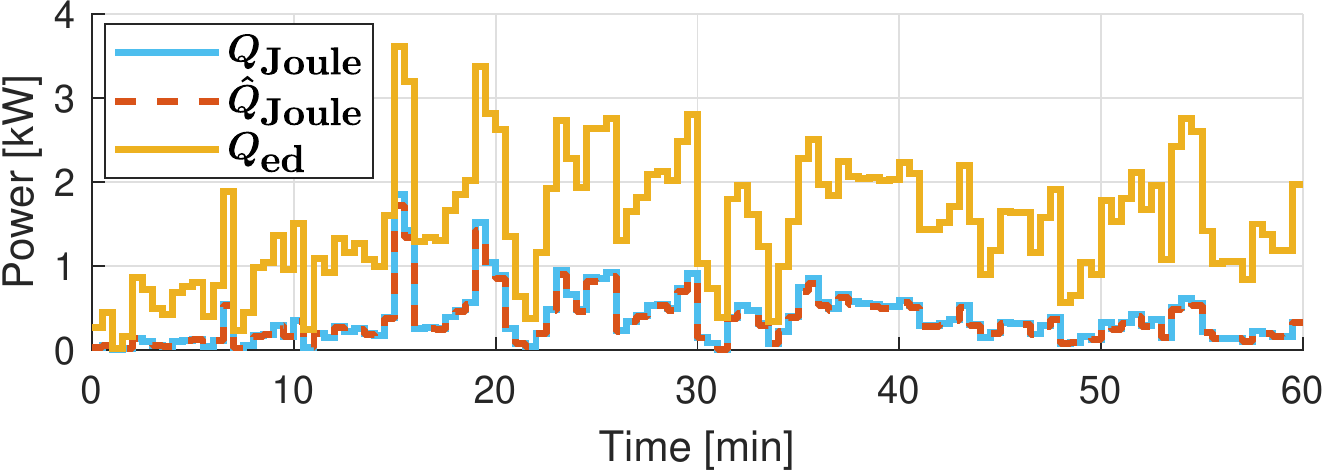}
\label{fig:juleed}
}
\subfigure[Leakage to ambient.]{

 \includegraphics[width=.9\linewidth]{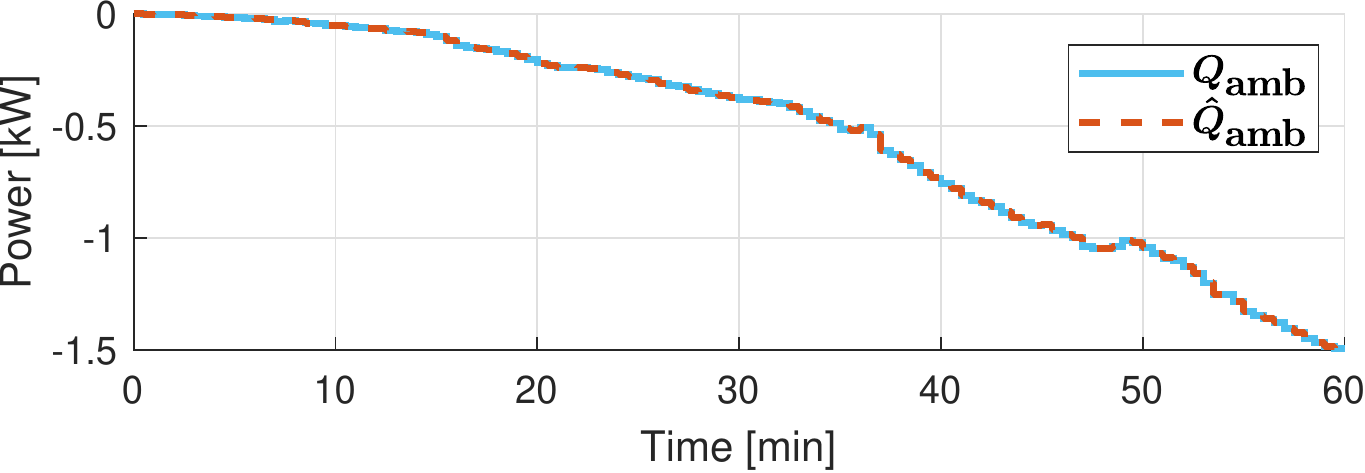}
\label{fig:amb}
}
\subfigure[Vehicle speed and battery electrical energy profiles.]{

 \includegraphics[width=.9\linewidth]{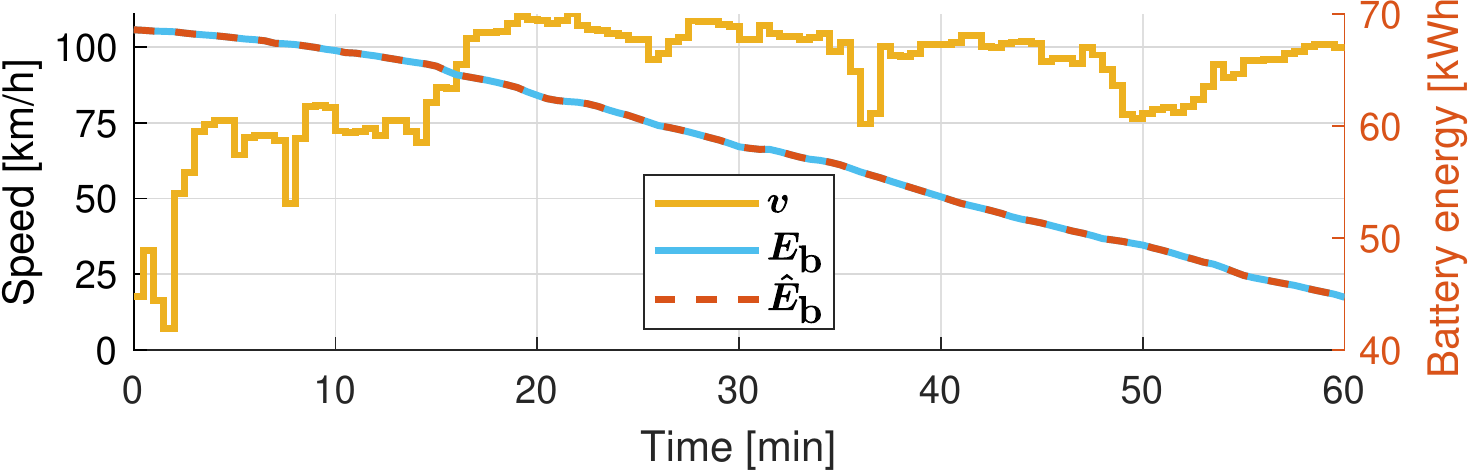}
\label{fig:vEb}
}
\caption{Comparison of the solutions obtained from solving the NLP and applying the heuristic approach.}\vspace{-.1cm}
\label{fig:res}
\end{figure}

\begin{table}
\begin{center}
\caption{Vehicle and Simulation Parameters}
\label{tab:par}
\begin{tabular}{l l}
\hline
Maximum batt. capacity & $C_\tx{p}=\SI{200}{Ah}$\vspace{0.05cm}\\ 
Time sampling interval & $\Delta t=\SI{30}{s}$\vspace{0.05cm}\\
EM max power & $\SI{350}{kW}$\vspace{0.05cm} \\
Max. battery power (discharging) & $P_\tx{b,dchg}^{\max}=\SI{350}{kW}$\vspace{0.05cm} \\
Min. battery power (charging) & $P_\tx{b,chg}^{\min}=\SI{-150}{kW}$\vspace{0.05cm} \\
Auxiliary load & $P_\tx{aux}=\SI{0.5}{kW}$\vspace{0.05cm} \\
HVCH power for heating cabin & $P^\tx{c}_{\tx{hvch}}=\SI{1.978}{kW}$ \vspace{0.05cm}\\
EM efficiency & $\eta_\tx{ed}^\tx{e}=\SI{90}{\%}$\vspace{0.05cm} \\
ED thermal efficiency & $\eta_\tx{ed}^\tx{Q}=\SI{80}{\%}$\vspace{0.05cm} \\
HVCH power to heat rate efficiency & $\eta_{\tx{hvch}}=\SI{87}{\%}$\vspace{0.05cm} \\
HVCH power to heat rate efficiency & $\eta_{\tx{hvch}}=\SI{87}{\%}$\vspace{0.05cm} \\
Initial battery temperature & $T_{\tx{b0}}=\SI{-7}{^\circ C}$\vspace{0.05cm} \\
Ambient temperature & $T_\tx{amb}=\SI{-7}{^\circ C}$\vspace{0.05cm} \\
Initial battery state of charge & $\tx{soc}_0=\SI{90}{\%}$\vspace{0.05cm} \\
Terminal battery state of charge & $\tx{soc}_\tx{f}=\SI{60}{\%}$\vspace{0.05cm} \\
\hline
\end{tabular}
\end{center}
\end{table}
In Fig~\ref{fig:res}, the obtained trajectories from solving the heuristic algorithm are illustrated together with the corresponding trajectories by solving the NLP \eqref{eq:p1}, where unsurprisingly the heuristic solution (states and control input) overlaps the NLP solution. The battery temperature increases significantly over the course of the trip, due to the Joule heating, ED circuit heating, and using HVCH. As mentioned earlier, the optimal solution shown in Fig.~\ref{fig:hvch} starts using the HVCH for battery heating \SI{28}{min} prior to arrival at the fast charging station, and leveling out the temperature at target temperature $\SI{25}{^\circ C}$ at the destination. The power corresponding to the thermal leakage to ambient in Fig.~\ref{fig:amb} is always non-positive as the battery temperature is never below the ambient temperature during the vehicle's trip. The gradual battery depletion in terms of SoC is also translated into total battery energy consumption, indicating that about $\SI{23}{kWh}$ is utilized throughout the trip. In Table II, the details of energy usage for the battery heating and leakage to ambient are reported for both the NLP and heuristic solutions. Note that the thermal energy due to the ED losses is the same value for both the NLP and heuristic solution, as the driving cycle is the same for both cases. The overall increase, i.e. \SI{1}{Wh}, in energy consumption is negligible for the investigated route.

\begin{table}
\caption{Energy Consumption [Wh]}\vspace{-0.1cm}
\label{tab:costvst}
\begin{center}
\begin{tabular}{ c|c|c } 
\textbf{Eng. component} & \textbf{NLP sol.}  & \textbf{Heuristic sol.}\\
\hline
\multirow{1}{10em}{Joule heating} & 361.4 & 345.1 \\
\multirow{1}{10em}{ED heating} & 1504 & 1504 \\
\multirow{1}{10em}{HVCH bat. heating} & 2019.2 & 2038.9 \\
\multirow{1}{10em}{Ambient leakage} & -526.1 & -525.8 \\
\multirow{1}{10em}{Total bat. eng.} & 23888 & 23889 \\
\end{tabular}
\end{center}
\end{table}

\section{Conclusion}\label{sec:con}
In this paper, the battery preheating has been addressed for a BEV driving in a cold weather towards a planned fast-charging station. To do so, a simple algorithm has been devised, by imitating the optimal battery preheating behaviour. According to the simulation results, the computational burden has been reduced significantly by applying the proposed approach, in the cost of only \SI{1}{Wh} increase in the total energy consumption. Such a fast strategy can be implemented on lower level control unit with limited performance, as the need for solving an NLP has been lifted.


\bibliography{bibliography_ah}

\end{document}